\documentclass[draft]{agujournal2019}
\usepackage{url} 
\usepackage{lineno}
\usepackage[inline]{trackchanges} 
\usepackage{soul}
\usepackage{amsmath}
\usepackage{multirow}
\usepackage{array}

\draftfalse

\journalname{Geophysical Research Letters}

\begin{document}

%
%

\title{Mesoscale soil moisture heterogeneity can locally amplify humid heat}

%
%

\authors{G Chagnaud\affil{1}, C M Taylor\affil{1,2}, L S Jackson\affil{3}, A Barber\affil{3, 4}, H Burns\affil{3, 4}, J H Marsham\affil{3,5}, and C E Birch\affil{3}}

\affiliation{1}{UK Centre for Ecology and Hydrology}
\affiliation{2}{National Centre for Earth Observation}
\affiliation{3}{School of Earth and Environment, University of Leeds}
\affiliation{4}{Centre for Environmental Modelling and Computation, University of Leeds}
\affiliation{5}{Met Office}

\correspondingauthor{G Chagnaud}{guicha@ceh.ac.uk}



\begin{keypoints}  
\item Humid heat associated with soil moisture heterogeneity is locally amplified by 1--4$^\circ$C as compared to uniform wet soil  
\item Subsidence due to soil moisture-induced mesoscale circulations more efficiently concentrates warm, humid air in a shallower boundary layer  
\item Background wind and wet-dry contrast magnitude control the relationship between soil moisture length-scale and humid heat amplification  
\end{keypoints}

\begin{abstract}  
Soil moisture is a key ingredient of humid heat through supplying moisture and modifying boundary layer properties. Soil moisture heterogeneity due to e.g., antecedent rainfall, can strongly influence weather patterns; yet, its effect on humid heat is poorly understood. Idealized numerical simulations are performed with a cloud-resolving ($\Delta x=$ 500 m), coupled land-atmosphere model wherein wet patches on length-scales $\lambda \in$ 25--150 km are prescribed. Compared to experiments with uniform soil moisture, humid heat is locally amplified by 1--4$^\circ$C, with maximum amplification for the critical soil moisture length-scale $\lambda_c=$ 50 km. Subsidence associated with a soil moisture-induced mesoscale circulation concentrates warm, humid air in a shallower boundary layer. The background wind and the magnitude of the wet-dry contrast control the relationship between $\lambda_c$ and the humid heat amplification. Based on observed soil moisture patterns, these results will help to predict extreme humid heat on city and county scales across the Tropics.
\end{abstract}

\section*{Plain Language Summary} 
Humidity exacerbates human heat stress by limiting the body’s cooling through reduced sweat evaporation. Humid heat extremes can occur when additional moisture is supplied in a warm environment, for instance from wetter soils following rainfall. The spatial variability of soil moisture is known to influence weather patterns, but its effects on humid heat are largely unexplored. To investigate how soil moisture heterogeneity influences humid heat, a series of simulations are performed with a high-resolution model of the coupled land-atmosphere system: circular wet patches of length 25--150 km across are imposed in the centre of an otherwise dry domain. Compared to a simulation with uniform soil moisture, humid heat is locally larger by 1 to 4$^\circ$C with soil moisture heterogeneity. Atmospheric circulations generated by soil moisture contrasts reduce the vertical mixing of air, leading to higher humid heat near the surface than would occur without soil moisture heterogeneity. Our results are robust across different metrics of humid heat and across various environments. They improve mechanistic understanding of humid heat, which will help to better predict hazardous humid heat hours to days in advance across vast swathes of the Tropics.


%
%

\section{Introduction}  \label{sec:introduction}

The habitats of many terrestrial species are within a few metres above the ground; their living conditions therefore depend largely on weather conditions near the surface, which involve land-atmosphere interactions on a wide range of time and spatial scales \cite{santanello_landatmosphere_2018}. Heatwaves, prolonged periods of high heat stress, are among the most detrimental weather phenomena to societies and ecosystems. The increase in the intensity, frequency, and duration of heatwaves as a result of global warming has already caused numerous excess deaths around the world and will continue to do so, even with moderate greenhouse gases emissions \cite{mitchell_attributing_2016, gasparrini_projections_2017, vicedo-cabrera_burden_2021, matthews_mortality_2025}.

Humidity is known to exacerbate human heat stress by limiting sweat evaporation. Humid heat can be quantified by various metrics, including wet-bulb temperature (Twb) and Heat Index (HI), and there is no consensus in the literature on which is more relevant to human health. \citeA{sherwood_adaptability_2010} suggest that an upper survivability limit for fit people is Twb=35$^\circ$C \cite<see also>{buzan_moist_2020}. Other studies show that adverse health effects can occur in vulnerable groups after $\sim$6 hours of exposure to Twb $\geq$ 24--25$^\circ$C \cite{vecellio_evaluating_2022, vanos_physiological_2023}, considerably increasing the prevalence of high Twb values as a global heat risk \cite{vecellio_greatly_2023}. The all-year 95$^\text{th}$ percentile of daily mean Twb (Twbd-95) is 24$^\circ$C on average in the Tropics and above 25$^\circ$C in many coastal and inland low-lying areas (Fig. S1a). In addition, there is strong seasonality in the timing of Twbd-95 (Fig. S1b), as dictated, for example, by the onset or cessation of rainy seasons \cite{ivanovich_subseasonal_2024, jackson_daily_2025}.

Recent studies have demonstrated that the highest Twb values are tied to large-scale atmospheric dynamics through stability and entrainment \cite{zhang_projections_2021, duan_moist_2024, li_atmospheric_2026}, but they also acknowledge that local processes can lead to extreme humid heat. Wet soils, driven, for instance, by antecedent rainfall or irrigation, support higher humid heat in southern Asia \cite{monteiro_characterization_2019, mishra_moist_2020}, the US \cite{krakauer_effect_2020, gurung_understanding_2024, chen_heat_2025}, China \cite{kang_north_2018, zou_agricultural_2024, yan_expanding_2025}, and across sub-Saharan Africa \cite{chagnaud_wet-bulb_2025}. Soil moisture (SM) can exert a strong control on near-surface humid heat through surface latent heat flux (LHF) being enhanced at the expense of surface sensible heat flux (SHF), which limits the growth of the boundary layer (BL) and concentrates more humid air in a shallower BL \cite{mishra_moist_2020, kong_regimes_2023, chagnaud_wet-bulb_2025, jackson_daily_2025}. This ``volume reduction" mechanism is especially efficient in regions/seasons where SM exerts a first-order control on the partitioning of available (radiative) energy, $R_{net}$, into surface heat fluxes. This typically occurs when i) insolation is abundant (i.e. cloud cover is minimal), and ii) soil moisture sits between the wilting point ($\theta_W$), below which soils are largely dry, and the critical point ($\theta_C$), above which soils are wet; the SM--evaporative fraction (EF = LHF / $R_{net}$) relationship then belongs to the water-limited regime and is said to be ``transitional'' \cite{seneviratne_investigating_2010, dirmeyer_terrestrial_2011}. The SM--EF regime is transitional in large swathes of sub-Saharan Africa, India, and Central America more than 50\% of the time during the humid heat season (Fig. S1c).

On daily timescales, soil moisture can exhibit spatial variability over distances ranging from a few hundred meters to several tens of kilometers. Spatial heterogeneity is a ubiquitous property of soil moisture in the Tropics (Fig. S1d). In the transitional SM--EF regime, spatial SHF contrasts resulting from soil moisture heterogeneity can trigger mesoscale circulations, in turn influencing the land surface--BL coupling. Having been extensively studied with a diversity of observational \cite{bhumralkar_observational_1973, mahrt_observations_1994, taylor_observational_2007, dixon_effect_2013, barton_case-study_2020}, analytical \cite{green_mesoscale_1980, dalu_vertical_1993, segal_nonclassical_1992} and numerical approaches \cite{ookouchi_evaluation_1984, pielke_nonlinear_1991, chen_impact_1994, wang_numerical_1998, patton_influence_2005, kang_effects_2008, van_heerwaarden_scaling_2014, han_largeeddy_2019, zhang_largeeddy_2023}, mesoscale circulations are enhanced for a range of spatial heterogeneity length-scales ($\lambda$). Furthermore, the large-scale wind, the amplitude of surface heterogeneity, and the thermodynamical structure of the atmosphere modulate the way mesoscale circulations form, grow, and decay.

The impacts of mesoscale circulations on land surface--BL coupling have been widely studied from the perspective of convection (see above references); yet, their effects on humid heat remain largely unexplored. Averaging across more than 200 heatwaves from a pan-African, convection-permitting model simulation, \citeA{chagnaud_wet-bulb_2025} found a mesoscale circulation and reduced BL height ($z_i$) in cases associated with positive soil moisture anomalies at $\lambda\approx$ 50 km. In these events, Twb was amplified by 0.5--0.6$^\circ$C compared to events associated with larger-scale soil moisture anomalies, where a mesoscale circulation was absent. This study suggests a pathway whereby soil moisture heterogeneity influences Twb extremes, calling for further investigation of the relationship between humid heat and soil moisture length-scale.

We examine how soil moisture heterogeneity influences near-surface humid heat by performing idealized simulations with a convection-resolving ($\Delta x=$ 500 m), coupled land-atmosphere model. The objectives are to determine i) whether a critical soil moisture length-scale $\lambda_c$ exists for which the effect on humid heat is maximized and, if so, ii) what are the mechanisms involved, and iii) how does $\lambda_c$ vary with environmental conditions and humid heat metrics. In addition to improving our understanding of the role of fine-scale land-atmosphere interactions in humid heat, providing answers to these questions would improve our capacity to use satellite observations of soil moisture --available even in regions of sparse weather station networks-- in combination with numerical weather models to predict humid heat extremes.

\section{Methods}   \label{sec:data_method}

\subsection{Idealized configuration of the Met Office Unified Model}

A series of simulations were run using an idealized configuration of the Met Office Unified Model (UM) coupled to the Joint UK Land Environment Simulator (JULES). The UM science configuration is based on version 3 of the Regional Atmosphere and Land configuration \cite{bush_third_2024}. The non-hydrostatic, fully compressible equations of motion are discretized on a 400 km $\times$ 400 km cartesian grid with 500 m horizontal grid spacing and solved with the ENDGame dynamical core \cite{wood_inherently_2014}, the time step of integration being 20 seconds. In the vertical, the model uses a Charney–Phillips grid \cite{charney_numerical_1953} with 90 levels, of which 67 are below 18 km, and a fixed model lid 40 km above the surface. Key parameterizations include radiation \cite{edwards_studies_1996}, microphysics \cite{field_implementation_2023}, and turbulence \cite{boutle_seamless_2014}. The surface--atmosphere exchanges of momentum, mass, and energy, together with the surface and sub-surface flows of water and heat are treated in JULES \cite{best_joint_2011}. Similar UM configurations have been used to study the atmospheric impacts of irrigation over India \cite{fletcher_effect_2022}, the effect of soil moisture variability on Sahelian storms \cite{maybee_homogeneous_2025}, and to test convection parameterization schemes \cite{lavender_use_2024}. The UM has also been run in a non-idealized configuration for climate and climate change studies at convection-permitting scales over Africa \cite{stratton_pan-african_2018, kendon_enhanced_2019, birch_future_2022}.

\subsection{Experimental setup and simulations}

To avoid contamination of soil moisture-induced circulations by flows due to topographic or vegetation heterogeneity, flat terrain and bare soil are used throughout the model domain. The hydraulic properties of bare soil are set to that of sand: the volumetric soil moisture content at wilting point and critical point are $\theta_W=$ 0.045 m$^3$ m$^{-3}$ and $\theta_C=$ 0.128 m$^3$ m$^{-3}$, respectively \cite{dharssi_new_2009}. The roughness lengths for momentum and scalars are set to $z_{0,\text{m}}=$ 10 cm and $z_{0,\text{h}}=$ 2 cm, corresponding to what would be found on tropical grassland of height $h=$ 1 m with $z_{0,\text{m}}$ parameterized as $z_{0,\text{m}}=h$/10 and $z_{0,\text{h}}$/$z_{0,\text{m}}=$ 0.2 \cite{best_joint_2011, stratton_pan-african_2018}. We use radiosonde observations of temperature and humidity from the African Monsoon Multidisciplinary Analysis field campaign \cite{parker_amma_2008}, with cyclic lateral boundary conditions mimicking an infinite domain and without large-scale advection of either quantity, as in \citeA{lavender_use_2024}. The background wind speed, $U$, is set to 4 m s$^{-1}$, consistent with the average large-scale wind speed found at lower atmospheric levels during heatwaves in \citeA{chagnaud_wet-bulb_2025}. $U$ flows from West to East and has a constant vertical profile. The simulations are initialized at midnight and run freely for 24 hours with the diurnal cycle of insolation of Niamey on 10 July, peaking at 950 W m$^{-2}$ at 1200 local time (LT). This configuration forms the reference setup.

We perform two control simulations with uniformly dry and uniformly wet soils respectively labeled UDRY and UWET. In UDRY, soil moisture is initialized at 0.3$\times\theta_C$ in the four soil layers. In UWET, soil moisture is initialized at $\theta_C$ everywhere. To reproduce soil moisture heterogeneity typically induced by rainfall or irrigation, we perform simulations where a soil moisture perturbation is prescribed as a circular wet patch positioned in the centre of the domain (see example in Fig. \ref{fig:fig1}): soil moisture is initialized at $\theta_C$ in the patch and 0.3$\times\theta_C$ elsewhere, in all soil layers. This soil moisture contrast ($\delta$) allows a direct comparison between perturbed and control simulations. The diameter of the patch, $\lambda$, varies from 25 to 150 km; each perturbed simulation is accordingly labeled P$\lambda$ (Table \ref{tab:experiments}). Restricting patch sizes to $\lambda \leq$ 150 km ensures that soil moisture-induced circulations remain within the limits of the domain. Model diagnostics are output as hourly means to minimise the impact of turbulence. Additional sets of UWET--P$\lambda$ simulations are performed to evaluate the sensitivity of the $\lambda$--humid heat relationship to $U$, $\delta$, the vertical profiles of potential temperature ($\gamma_{\theta}$) and humidity ($\gamma_{q}$) (Table \ref{tab:experiments} and Text S2).

Examining BL properties and surface energy budget terms in control simulations shows that, in the range of soil moisture values utilized here, the BL is very sensitive to the land surface (Fig. S2 and Text S3), as summarized by 1200 LT SHF and 1500 LT $z_i$ values that are more than double in UDRY compared to UWET (Table \ref{tab:experiments}). Since cloud cover remains low (Fig. S3) and no significant rainfall occurs (not shown), this land surface--BL coupling regime remains similar throughout the simulations.

\begin{table}[h!]
\centering
\begin{tabular}{l|l|l|l|l|l|l|l|l|}
\parbox{0.1cm}{ } & \parbox{1cm}{Name}  & \parbox{2.15cm}{$\lambda$\par[km]} & \parbox{0.7cm}{SM\par[\%$\theta_C$]} & \parbox{1.6cm}{$\gamma_{\theta}$\par[$^\circ$C km$^{-1}$]} & \parbox{1.15cm}{$\gamma_{q}$\par[g kg$^{-1}$ km$^{-1}$]} & \parbox{1.125cm}{$U$\par[m s$^{-1}$]} & \parbox{1.5cm}{SHF(1200)\par[W m$^{-2}$]} & \parbox{1.1cm}{$z_i$(1500)\par[m]} \tabularnewline[3mm] \hline
\multirow{4}{*}{\rotatebox[origin=l]{90}{REF}}         & UDRY       &   -         &  30  & 4.3     & -5.1 & 4 &  460 & 3470  \\
                                                       & UWET       &   -         &  100 & 4.3     & -5.1 & 4 &  163 & 1678 \\
                                                       & \multirow{3}{*}{P$\lambda$} & \parbox{2.25cm}{25, 30, 35, 50, 75, 100, 125, 150} & 30--100 & 4.3   & -5.1 & 4 &  -   &  - \\ \hline\
\multirow{8}{*}{\rotatebox[origin=l]{90}{SENSITIVITY}} & $U-$       & 15, 25, 35, 50, 100 & 30--100 & 4.3      & -5.1 & 2 & 160  & 1657  \\
                                                       & $U$+       & 25, 50, 100, 150 & 30--100 & 4.3 & -5.1 & 8 & 170  & 1696  \\
                                                       & $\delta-$  & 25, 50, 100 & 65--100 & 4.3      & -5.1 & 4 & 163 & 1678 \\
                                                       & $\delta$+  & 25, 50, 100 &  30--170 & 4.3     & -5.1 & 4 & 77  & 1220 \\
                                                       & $\gamma_{\theta}-$ & 25, 50, 100 & 30--100 & 3.5 & -5.1 & 4 & 153 & 1945 \\
                                                       & $\gamma_{\theta}$+ & 25, 50, 100 & 30--100 & 5.1 & -5.1 & 4 & 180 & 1387 \\
                                                       & $\gamma_{q}-$ & 25, 50, 100 &  30--100 & 4.3  & -3.8 & 4 & 165 & 1500 \\
                                                       & $\gamma_{q}+$ & 25, 50, 100 &  30--100 & 4.3  & -6.4 & 4 & 139  & 1828   \\  \hline
\end{tabular}
\caption{Experiment names and main characteristics for the reference setup (REF) and the sensitivity experiments (SENSITIVITY): $\lambda$ is the wet patch diameter; SM is the volumetric soil moisture content in all soil layers (in \% of $\theta_C$, for perturbed runs the values are displayed as outside--inside patch); $\gamma_{\theta}$ and $\gamma_{q}$ are respectively the vertical gradients of potential temperature and specific humidity used as boundary conditions (both averaged between the surface and 10 km); $U$ is the background wind speed; SHF(1200) and $z_i$(1500) are the domain-averaged sensible heat flux at 1200 LT and boundary layer height at 1500 LT, respectively. SHF(1200) and $z_i$(1500) values for the SENSITIVITY setups are from UWET simulations.}
\label{tab:experiments}
\end{table}

\subsection{Calculation of humid heat metrics}

Wet-bulb temperature is calculated with the Davies-Jones formula \cite{davies-jones_efficient_2008} using hourly near-surface dry-bulb temperature and specific humidity, and hourly surface pressure. The Heat Index (HI) is calculated following Rothfusz's polynomial approximation of a human thermoregulation model \cite{steadman_assessment_1979}, with a correction applied following \citeA{romps_chronically_2022}. We conduct our analysis using Twb and compare the results to HI.

\section{Results and discussion}   \label{sec:results}

\subsection{Near-surface atmospheric state in simulations with uniform and heterogeneous soil moisture}

In the simulation with the 50 km patch (P50), cooler and moister air is found above the wet patch compared to the surroundings (Fig. \ref{fig:fig1}a and b). The net effect on Twb and HI is an amplification of 2$^\circ$C and 1$^\circ$C, respectively, over the wet patch compared to the domain average (Fig. \ref{fig:fig1}c and Fig. S4a). It is noteworthy that Twb is 2 to 4$^\circ$C warmer in UWET compared to UDRY, showing how humid heat is enhanced over wet soil (Fig. \ref{fig:fig1}c, inset). The 0800--1900 LT average $z_i$ is 580 m over the wet patch, whereas the domain average is 1530 m; for reference, the daytime average $z_i$ is 1555 m in UDRY and 800 m in UWET (Fig. \ref{fig:fig1}d, inset). The 10-meter wind is slower upwind of the wet patch and faster downwind, compared to uniform soil moisture experiments (Fig. \ref{fig:fig1}c).

The four variables also display noticeable sub-daily variability. The wet patch mean Twb has a first peak at 0800 LT and a second, higher peak at 1900 LT (Fig. \ref{fig:fig1}c, inset), a bi-modal diurnal cycle typical of semi-arid regions \cite{birch_future_2022}. Twb values closely follow the diurnal evolution of $q$ (Fig. \ref{fig:fig1}b, inset), and the mid-afternoon dip in Twb and $q$ corresponds to the peak in $T$ and $z_i$. The highest Twb value found at 1900 LT over the wet patch is associated with values of $T$ that are slightly larger than found over uniformly wet soil, alongside larger $q$ and smaller $z_i$. This analysis highlights the interplay between $T$, $q$, and $z_i$ in modulating the space-time structure of Twb. Importantly, Twb and HI are larger over locally wetter soils than found with uniform soil moisture, in association with warmer and more humid air mixed in a shallower BL. Next, we investigate how this local humid heat amplification varies with the length-scale of soil moisture heterogeneity ($\lambda$).


\begin{figure}[ht!]
\centering
\includegraphics[width=\textwidth]{fig1.pdf}
\caption{Daily mean (a) $T$, (b) $q$, (c) Twb, and (d) $z_i$ in P50. In (c) the daily mean 10-meter wind is shown with arrows. Insets show spatially averaged evolution of 3-hourly values: green, blue and black lines are domain averages for UDRY, UWET, and P50, respectively; orange line is wet patch average in P50 (see legend). The dashed green circle in (c), with radius $r=$10 km, indicates an area used in Section \ref{sec:results}.2.}
\label{fig:fig1}
\end{figure}

\subsection{Local humid heat amplification by mesoscale soil moisture heterogeneity}

\begin{figure}[ht!]
\centering
\includegraphics[width=\textwidth]{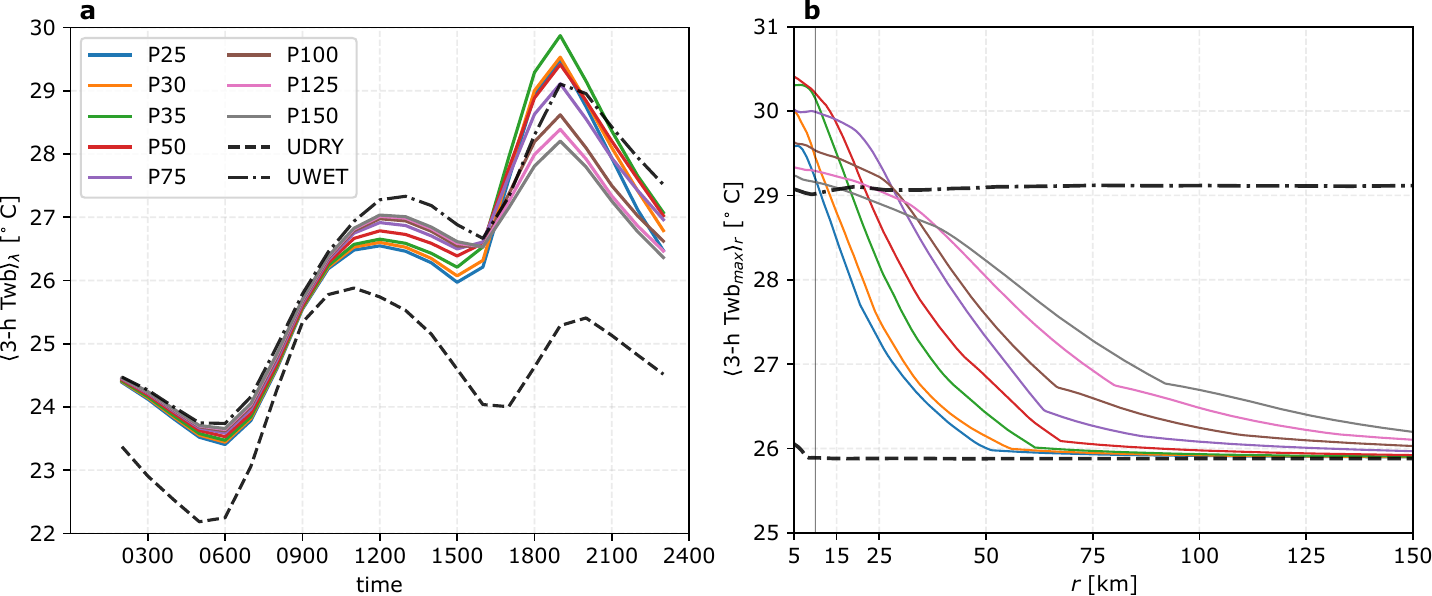}
\caption{(a) Diurnal cycle of 3-hourly Twb in perturbed experiments (color lines, see legend), UDRY (dashed line), and UWET (dotted-dashed line). In P$\lambda$, wet patch averages are considered whereas in UDRY and UWET the domain-average Twb is plotted. (b) Maximum 3-hourly Twb values averaged over circular areas centered at x$_0$=200 + $\lambda$/4 km and of radius $r$ (in km, horizontal axis) for all perturbed experiments and UWET (see legend in (a)).}
\label{fig:fig2}
\end{figure}

The diurnal cycle of 3-hourly Twb values peaks at 1900 LT for all wet soil experiments (Fig. \ref{fig:fig2}a). In addition, wet patch averages in P$\lambda$ are larger than the domain average in UWET when $\lambda \leq$ 50 km. Similar pattern is found for HI with a peak at 1800 LT and $\lambda \leq 75$ km (Fig. S4b). Hence, peak humid heat values over smaller-scale soil moisture heterogeneity are amplified compared to those over larger-scale wet features.

There is substantial within-patch spatial variability in peak humid heat: visual inspection of Twb fields at 1900 LT reveals that the highest Twb values systematically occur over the downwind half of the patch, that is, between the centre of the domain (x$_0$) and the right-hand edge of the patch (Fig. S5). We therefore plot, for each experiment, the 3-hourly Twb averaged over circular areas with radius $r \in$ [5--150 km] centered on x=x$_0$ + $\lambda/4$, denoted $\langle$Twb$_{max}\rangle_{r}$ (see an example in Fig. \ref{fig:fig1}c for $r=10$ km). In UDRY and UWET, $\langle$Twb$_{max}\rangle_{r}$ is constant with $r$, reflecting spatial homogeneity (Fig. \ref{fig:fig2}b). In perturbed experiments, $\langle$Twb$_{max}\rangle_{r}$ has a much stronger dependence on $r$. Notably, $\langle$Twb$_{max}\rangle_{r}$ in perturbed experiments is 0.2--1.2$^\circ$C larger than in UWET for $r \leq$ 10 km, for all $\lambda$. This small-scale Twb amplification is largest for $\lambda=$ 50 km and reduces for smaller and larger wet patches. A similar behaviour is found for the Heat Index, with $\langle$HI$_{max}\rangle_{r}$ up to 4$^\circ$C larger in perturbed than in control experiments for $r < 20$ km and $\lambda \leq$ 100 km (Figs. S6 and S7). However, locally higher HI values can be found at other times of day and in different locations (not shown), likely involving other SM-related mechanisms due to the weaker control of humidity on HI compared to Twb \cite{sherwood_how_2018}.

This analysis shows that in the experimental setup utilized here, a critical soil moisture length-scale $\lambda_c$ exists for which humid heat is maximized on length-scales up to 10--20 km across (300--1300 km$^2$). This area of locally larger Twb/HI values is observed consistently in all perturbed experiments, implying a deterministic cause, in contrast to turbulence-induced pockets of slightly higher humid heat values randomly found in control runs. In the next section, we investigate the land surface--BL coupling with a view to shedding light on the causal chain involved in this local humid heat amplification.


\subsection{Soil moisture-induced mesoscale circulation control on boundary layer dynamics}

\begin{figure}[ht!]
\centering
\includegraphics[width=\textwidth]{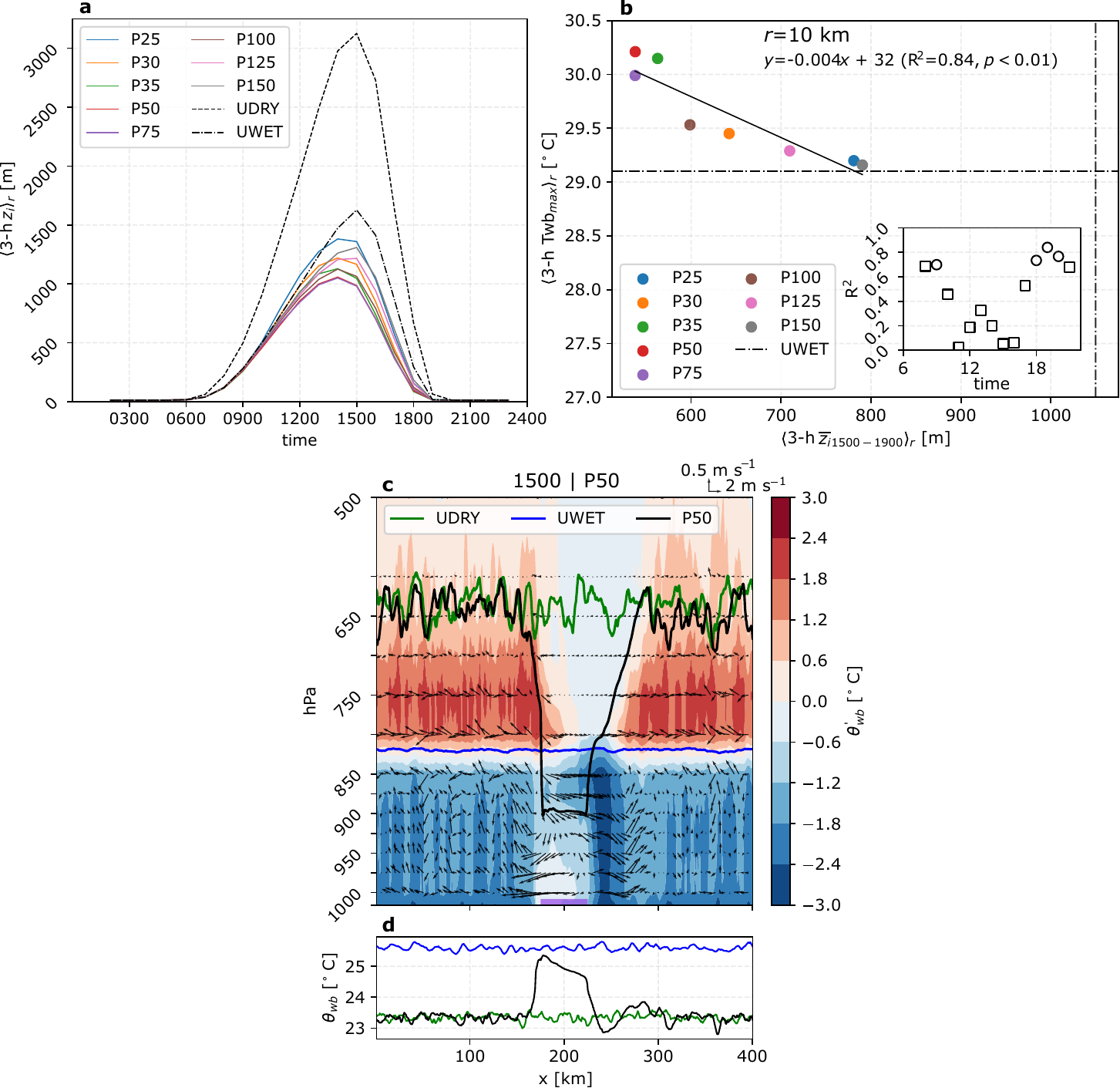}
\caption{(a) Diurnal cycle of 3-hourly $z_i$ in perturbed experiments (color lines, see legend), UDRY (dashed line), and UWET (dotted-dashed line). In P$\lambda$, $z_i$ is averaged over the area of maximum Twb ($r=$ 10 km), whereas in UDRY and UWET the domain average is plotted. (b) 3-hourly Twb at 1900 LT as a function of 1500--1900 LT mean 3-hourly BL height in perturbed experiments (dots); both quantities are averaged over a circular area of radius $r$=10 km centered on x=x$_0$ + $\lambda/4$, as inferred in the previous section. The corresponding domain-mean values in UWET are shown with dotted-dashed lines. The inset in the bottom right corner shows the R$^2$ of the linear regression between $\langle$Twb$_{max}\rangle_{10}$ sampled at each daytime hour and $\overline{z_i}$ averaged over the 4 preceding hours; significant regressions (at the 1\% level) are indicated with circles. (c) Vertical cross section in the longitudinal direction at 1500 LT in P50. Potential wet-bulb temperature anomaly ($\theta_{wb}'$, in $^\circ$C) is shown with shading and wind anomaly (in m s$^{-1}$) with arrows (anomalies are calculated as P50-UWET). Solid lines denote $z_i$ in UDRY (green), UWET (blue), and P50 (black). (d) Near-surface $\theta_{wb}$ at 1500 LT, with colors similar to the legend in (a). Longitudinal cross section values (panels c and d) are averaged along a 20 km band centered at y$_0$=200 km.}
\label{fig:fig3}
\end{figure}

In all perturbed experiments, the BL over the area of maximum Twb is shallower than found in their uniform soil moisture counterparts (Fig. \ref{fig:fig3}a). This suppressed BL development is especially prominent from noon through to the evening BL collapse. Figure \ref{fig:fig3}b, which shows for each perturbed experiment, $\langle$Twb$_{max}\rangle_{r=10\;\text{km}}$ as a function of the 1500--1900 LT average $z_i$, establishes a significant relationship (at the 1\% level) between afternoon $z_i$ and evening (peak) Twb (this result is robust to the choice of afternoon window). The inset in Fig. \ref{fig:fig3}b shows the strength of the statistical link between Twb sampled at each daytime hour and $z_i$ averaged over the 4 preceding hours: the $z_i$--Twb relationship is maximized in the late afternoon-evening. This link also holds between afternoon $z_i$ and evening HI, with a linear relationship significant at the 5\% level (Fig. S8).

Investigating BL dynamics in response to soil moisture heterogeneity provides process-based insight into the $z_i$--Twb relationship. Surface turbulent fluxes peak at 1200 LT (Fig. S2b and d), but the circulation is more intense a few hours later, when the balance between the near-surface pressure gradient that grows in response to the thermal contrast, on the one hand, and the dilution of the thermal contrast through horizontal advection and mixing, on the other hand, is optimal \cite<e.g.,>{segal_nonclassical_1992}. Given the clear relationship between afternoon $z_i$ and evening humid heat amplification, we examine BL dynamics during the afternoon.

A well-organized overturning circulation develops in perturbed experiments, as shown with the vertical cross section of potential wet-bulb temperature ($\theta_{wb}$) and wind anomalies at 1500 LT (Fig. \ref{fig:fig3}c). The mesoscale horizontal flow in the longitudinal direction reaches 4--5 m s$^{-1}$ in the lowest atmospheric layers, on both sides of the wet patch; this flow opposes and deflects the background wind on the upwind edge of the patch (Fig. \ref{fig:fig1}c), where narrow updrafts develop due to mass conservation. More diffuse downdrafts are located above the wet patch and further downstream. At this time of day, the BL $\theta_{wb}$ is 2--3$^\circ$C cooler in P50 than in UWET except above the wet patch, where this anomaly reduces to $\approx-1^\circ$C (Fig. \ref{fig:fig3}c and d) before turning positive from 1700 LT onward (Fig. \ref{fig:fig2}a). In P50, $z_i$ is similar to UDRY over dry areas and, most importantly, lower than in UWET over the wet patch. The subsiding branch of the mesoscale circulation therefore suppresses BL growth above wet soil, as seen in Figure \ref{fig:fig1}d and consistent with previous studies \cite{heerwaarden_relative_2008, suhring_effect_2014}. Over the wet patch, cloud cover change is limited to $\pm$10\% (Fig. S9a), resulting in a $R_{net}$ relative change $<$1\% (Fig. S9b). The effect of vertical mesoscale heat flux on the near-surface moist enthalpy being likely small (Fig. \ref{fig:fig3}c), combined with very little change in $R_{net}$ compared to changes in $z_i$ (Fig. S10), lead to the conclusion that BL growth reduction in response to soil moisture heterogeneity is instrumental in amplifying humid heat.


\subsection{Dependence to environmental conditions}

\begin{figure}[ht!]
\centering
\includegraphics[width=\textwidth]{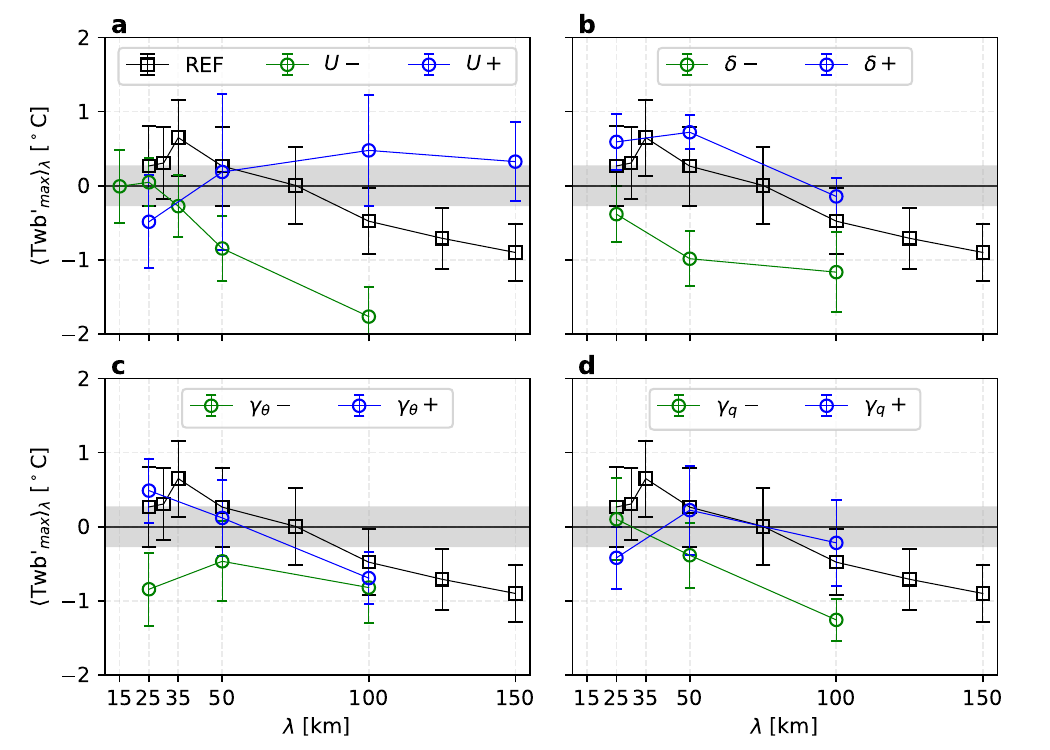}
\caption{Wet patch averaged maximum Twb anomaly ($\langle \text{Twb'}_{max}\rangle _{\lambda}$, Twb' = Twb$_{\text{P}\lambda}$-$\text{Twb}_{\text{UWET}}$; y-axis) for different values of soil moisture length-scale ($\lambda$; x-axis) in various sensitivity experiments (see Table 1): (a) $U-$ and $U+$, (b) $\delta-$ and $\delta+$, (c) $\gamma_{\theta}-$ and $\gamma_{\theta}+$, and (d) $\gamma_{q}-$ and $\gamma_{q}+$. Within-patch spatial variability is shown with error bars corresponding to $\pm\sigma$ about the spatial mean. Average spatial variability across UWET experiments is shown with grey shading corresponding to $\pm\sigma$ about the zero line.}
\label{fig:fig4}
\end{figure}

The relationship between soil moisture heterogeneity, boundary layer height, and humid heat involves an array of processes whose relative contributions to changes in the land surface--BL coupling depend on larger-scale environmental conditions. In sensitivity simulations (Table \ref{tab:experiments} and Text S2), Twb maxima occur at various locations within the wet patch (Fig. S11). To provide a fair comparison across all sensitivity experiments, wet patch averages of Twb maximum anomalies ($\langle\text{Twb'}_{max}\rangle_{\lambda}$) are considered hereafter, with the limitation that the \textit{true} Twb maxima may be underestimated. $\langle\text{Twb'}_{max}\rangle_{\lambda}$ is calculated as the difference between perturbed and UWET values. $\langle\text{Twb'}_{max}\rangle_{\lambda}$ values for the reference setup (`REF' in Table 1) are also indicated, noting that the maximum value is found for $\lambda=$ 35 km (Fig. \ref{fig:fig4}).

Mesoscale circulations are sensitive to background wind speed, with larger patch sizes required to generate a circulation with higher $U$, unless the magnitude of the surface heterogeneity ($\delta$) also increases \cite{segal_nonclassical_1992}. The relationship between $\lambda$ and $\langle\text{Twb'}_{max}\rangle_{\lambda}$ is consistent with this theoretical expectation: in $U+$, the critical soil moisture length-scale ($\lambda_c$) is shifted towards larger values and the Twb amplification is of similar magnitude as in REF, but covers a wider range of soil moisture length-scales (Fig. \ref{fig:fig4}a). By contrast, in $U-$, $\lambda_c$ is smaller and $\langle\text{Twb'}_{max}\rangle_{\lambda}$ values are reduced. Increased soil moisture contrast ($\delta+$) leads to larger Twb amplification compared to REF, whereas decreased soil moisture contrast ($\delta-$) results in negative $\langle\text{Twb'}_{max}\rangle_{\lambda}$ (Fig. \ref{fig:fig4}b).

The sensitivity of BL properties to soil moisture states also depends on the thermodynamical structure of the atmosphere \cite<e.g.,>{findell_atmospheric_2003}. With decreased atmospheric stability ($\gamma_{\theta}-$) the shape of the $\lambda$--$\langle\text{Twb'}_{max}\rangle_{\lambda}$ relationship is preserved but $\langle\text{Twb'}_{max}\rangle_{\lambda}$ values are negative (Fig. \ref{fig:fig4}c). Increased stability ($\gamma_{\theta}+$) yields very similar results to REF, although no critical $\lambda$ is found based on the limited number of soil moisture length-scales considered. With increased vertical humidity gradient ($\gamma_{q}+$), $\lambda_c\approx50$ km and $\langle\text{Twb'}_{max}\rangle_{\lambda=50}$ is equal to the REF value (Fig. \ref{fig:fig4}d). With reduced vertical humidity gradient ($\gamma_{q}-$), $\langle\text{Twb'}_{max}\rangle_{\lambda}$ values decrease with increasing $\lambda$ and are negative for $\lambda>$ 25 km.

Qualitatively similar conclusions can be drawn for HI, a notable exception being the HI amplification in $U-$ for $\lambda \leq 35$ km (Fig. S12). Therefore, the behaviour of the relationship between soil moisture length-scale and humid heat in various environments is consistent across a range of humid heat metrics. Other metrics consider the role of wind and radiation in human thermoregulation. In this respect, it should be noted that subsidence can affect both cloud cover and wind speed locally. The net effect of cloud cover reduction --involving changes in longwave and shortwave radiations-- on these metrics is difficult to guess, but weaker wind would reduce sweat evaporation rates, enhancing heat stress.

Three additional points should be noted: i) due to computational limitations, the full range of soil moisture length-scales, for each setup, was not explored, and $\lambda_c$ in these sensitivity runs may be considered cautiously; ii) the effects of each parameter have been examined separately, and additional simulations are needed to better document the combined effects of $U$ and $\delta$, alongside the thermodynamical structure of the atmosphere, on the land surface--BL coupling and their consequences for humid heat amplification (or lack thereof); iii) the sensitivity to other parameters, such as heterogeneity sharpness and land use, especially to represent urban areas, should be considered in future work.


\section{Conclusion}  \label{sec:conclusion}

A series of simulations performed with an idealized configuration of a convection-resolving ($\Delta x=$ 500 m), coupled land-atmosphere model demonstrates that mesoscale soil moisture heterogeneity can amplify humid heat by 1 to 4$^\circ$C on scales of 10--20 km across compared to uniform soil moisture. A critical soil moisture length-scale $\lambda_c=$ 50 km was found, with the background wind speed, the magnitude of the wet--dry contrast, and the vertical profiles of temperature and humidity determining the relationship between $\lambda_c$ and humid heat amplification. The most extreme humid heat values are thus more likely over locally wetter soils, as found in \citeA{chagnaud_wet-bulb_2025} based on a free-running, convective-scale climate model simulation. Numerical weather and climate models that do not represent soil moisture heterogeneity on the key spatial scales of 10 to 100 km will not simulate the associated mesoscale circulations, nor the humid heat maxima. Such circulations and maxima are also difficult to observe \cite{bou-zeid_persistent_2020}.

Our results point to the possibility of predicting extreme humid heat on city and county scales in regions of strong land-atmosphere coupling, found in the Tropics and beyond \cite{koster_regions_2004, kong_regimes_2023}, based on observed soil moisture patterns. Several tropical regions are climatologically close to hazardous physiological thresholds \cite{matthews_humid_2025}, meaning high quality early warning is essential. This issue is becoming especially pressing as global warming increases temperature and humidity worldwide and shifts environments from energy- to water-limited regimes \cite{denissen_widespread_2022, hsu_soil_2023, coffel_recent_2024}.


\section*{Open Research Section}
All simulations were run on ARCHER2, the UK's National Supercomputing Service \cite{beckett_archer2_2024}. The simulation outputs are available upon request from the authors on the H2X repository of JASMIN, the UK’s collaborative data analysis environment (\url{https://www.jasmin.ac.uk}). ERA5 data are available from the Copernicus Climate Change Service (C3S) Climate Data Store (CDS) at \citeA{hersbach_era5_2023}. The code used to calculate the wet-bulb temperature is available at \url{https://github.com/cr2630git/wetbulb_dj08_spedup}. The authors thank the Python community for developing and managing the xarray and matplotlib Python libraries, among many others. The codes used to process the data and plot the figures are available at \citeA{chagnaud_guichagh2x_2026}.

\section*{Conflict of Interest declaration}
The authors declare there are no conflicts of interest for this manuscript.

\acknowledgments
The authors thank the two reviewers for their insightful feedback. This work was supported by the NERC (grant numbers NE/X013618/1 and NE/X013596/1).

\nocite{hersbach_era5_2020, davies-jones_efficient_2008, romps_chronically_2022, schwingshackl_quantifying_2017}


%
\bibliography{mybib}
%

\end{document}